\documentclass[doublecol,aps,epl,preprintnumbers,nofootinbib,floatfix]{epl2}
\usepackage{graphicx,color,amsmath}
\usepackage{caption,subcaption}
\usepackage{amsfonts}
\usepackage{ae}
\usepackage{amsmath}
\usepackage[english]{babel}

\def\bnabla{\mbox{\boldmath $\nabla $}}

\def\CS{Carnahan-Starling}

\begin{document}

\title{General theory of asymmetric steric interactions in
  electrostatic double layers} \author{A. C. Maggs \inst{1} \and
  R. Podgornik \inst{2}} \shortauthor{Maggs and Podgornik}
\shorttitle{Steric effects in electrostatic double layers}

\institute{\inst{1} Physico-chimie th\'eorique, Gulliver, ESPCI-CNRS,
  10 rue Vauquelin 75005 Paris, France \\ \inst{2} Department of
  Theoretical Physics, J. Stefan Institute and Department of Physics,
  Faculty of Mathematics and Physics, University of Ljubljana, SI-1000
  Ljubljana, Slovenia}

 \date{}
  
\abstract{We study the Poisson-Boltzmann equation in the context of
  dense charged fluids where steric effects become important. We
  generalise the lattice gas theory by introducing a Flory-Huggins
  entropy for ions of differing volumes and then compare the effective
  free energy density to other approximations, valid for more
  realistic equations of state, such as the {\CS} approximation and
  find strong differences in the shapes of the free energy
  functions. {We solve the {\CS} model in the high density limit,
    and demonstrate a slow, power-law convergence at high
    potentials}. We elucidate how equivalent convex free energy functions
  can be constructed that describe steric effects in a manner which is
  more convenient for numerical minimisation.}

\maketitle
  
\section{Introduction}

\begin{figure}
\begin{center}
  \includegraphics[scale=0.17]{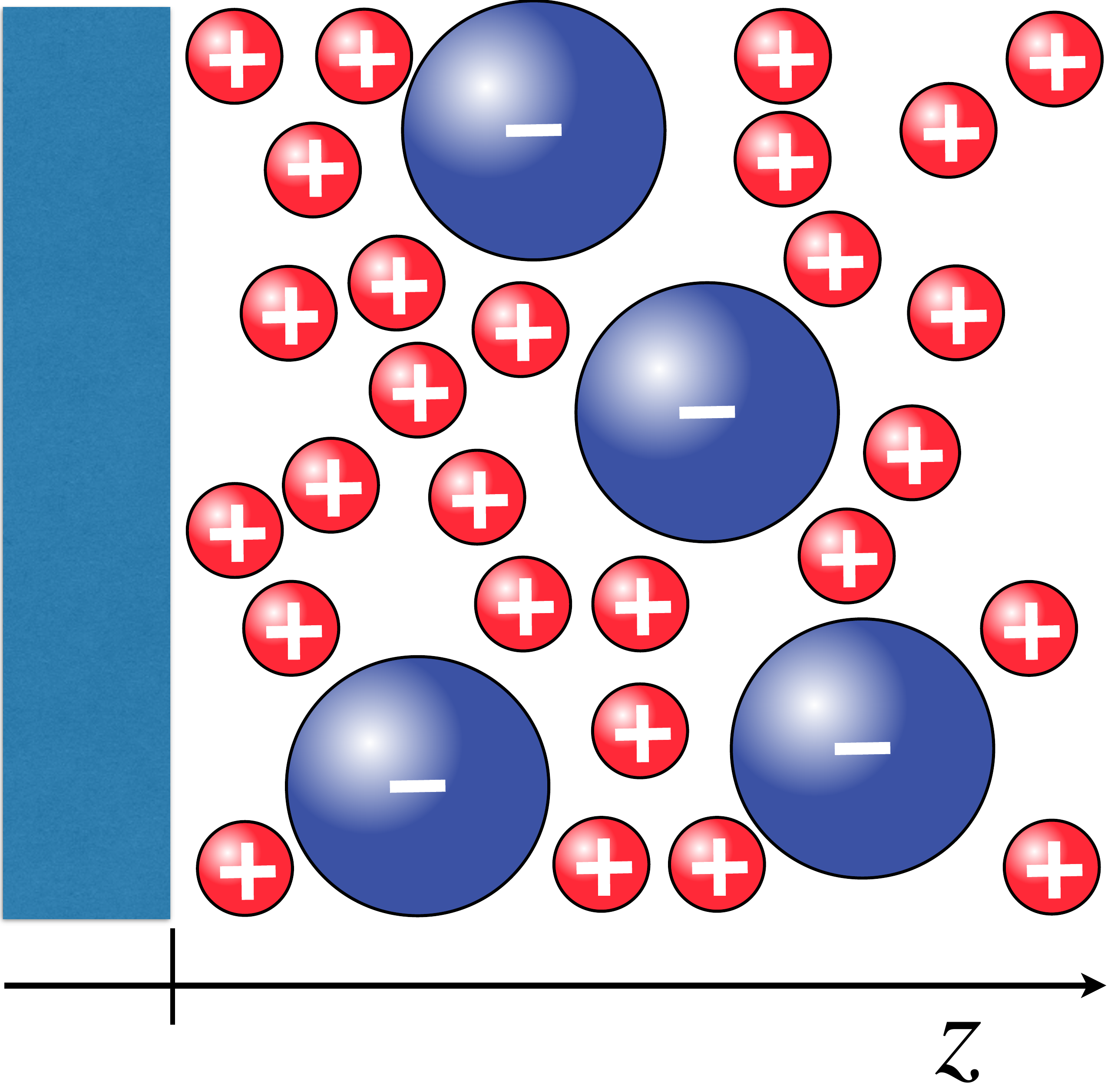}
  \caption{\label{fig:sch} A schematic representation of the asymmetric system composed of particles with unequal sizes and unequal valences in the vicinity of a charged surface. }
\end{center}  
\end{figure}

In the theory of ionic liquids \cite{Kornyshev-Fedorov} steric effects
are of particular importance since the packing of ions can be
especially dense \cite{Bazant}. The most common and simplest analytic approach to
these effects is via the {\em lattice gas mean-field
approximation} \cite{Kornyshev-rev}. This methodology can be furthermore extended
to a general local thermodynamic approach for any model of
inhomogeneous fluids \cite{Widom}. In this way one can connect the
equation of state for any reference uncharged fluid, not only a
lattice gas, with a full description of the same fluid with charged
particles on the mean-field electrostatics level, generalizing in this
way the Poisson-Boltzmann theory with consistent inclusion of packing
effects. This approach is particularly relevant for analysis of dense
electric double layers as arise in the context of ionic liquids or
dense Coulomb fluids in general \cite{Biesheuvel}. We will use this
{\em general local thermodynamics approach} in conjunction with two such
model equations of state: the asymmetric lattice gas approximation and
the asymmetric Carnahan-Starling approximation. The size asymmetry as
well as the charge asymmetry, Fig. \ref{fig:sch}, that this approach allows us to analyse,
are fundamentally important for understanding the nature of the
electrostatic double layers.

In what follows we will first formulate the local thermodynamics
mean-field approach to Coulomb fluids and then apply it to the
asymmetric lattice gas, derived within the Flory-Huggins lattice
approximation, comparing its results with the asymmetric
Carnahan-Starling approximation. As a sideline we also derive several
useful general relations valid specifically for the asymmetric lattice
gas approximation in the context of electrostatic double layers.

\section{General formulation}

We proceed by studying the Legendre transform of the free energy
density $f(c_1, c_2)$ of an isothermal ($T = const$) binary mixture
\begin{equation}
  f(c_1, c_2) - \mu_1 c_1 - \mu_2 c_2,
  \label{dapois}
\end{equation}
where $c_{1,2}$ are the densities of the two components, and the chemical potentials $\mu_{1,2}$ are
defined as
\begin{equation}
  \mu_{1,2} = \frac{\partial f(c_1, c_2)}{\partial c_{1,2}}.
  \label{ekrw}
\end{equation}
By the well known thermodynamic relationships \cite{Widom} the
Legendre transform eq.~(\ref{dapois}) equals
\begin{equation}
  f(c_1, c_2) - \frac{\partial f(c_1, c_2)}{\partial c_{1}} c_1 -\frac{\partial f(c_1, c_2)}{\partial c_{2}} c_2 = - p(c_1, c_2), 
\end{equation}
where $p(c_1, c_2)$ is the thermodynamic pressure, or the equation of
state. For the inhomogeneous case we now invoke the local thermodynamic
approximation so that the inhomogeneity is described solely via the
coordinate dependence of the densities, but the form of the
thermodynamic potential remains the same as in the bulk,
\begin{equation} {\cal F} = \int_V\!\!\!d^3{\bf r} \left( f(c_1, c_2)
    - \mu_1 c_1 - \mu_2 c_2\right) = - \int_V\!\!\!d^3{\bf r}~p(\mu_1,
  \mu_2).
  \label{dfgwy}
\end{equation}
In the case of charged particles one needs to consider also the
electrostatic energy and its coupling to the density of the particles
via the Poisson equation, on top of the reference free energy of
uncharged particles. The corresponding thermodynamic potential of the charged binary mixture then
assumes the form
\begin{align}
  {\cal F}[c_1, c_2, {\bf D} ] = \int_V\!\!\!d^3{\bf r} \left( f(c_1, c_2) - \mu_1 c_1 - \mu_2 c_2\right)  + \nonumber\\
  + \int_V\!\!\!d^3{\bf r} \left( \frac{{\bf D}^2}{2\varepsilon} -
  \psi\left( \bnabla\cdot {\bf D} - e (z_1 c_1 - z_2
  c_2)\right)\right),
  \label{cdrtqwy}
\end{align}
where ${\bf D} = {\bf D}({\bf r})$ is the dielectric displacement
field, $\varepsilon = \epsilon\epsilon_0$ with $\epsilon$ the relative dielectric permittivity, $z_{1,2}$ are the valencies of the two charged species and
$\psi = \psi({\bf r})$ is now the Lagrange multiplier field that
ensures the local imposition of Gauss' law \cite{localpb}.  We can
write this expression in an alternative form as
\begin{align}
  {\cal F} =&  \int_V\!\!\!d^3{\bf r} \left( f(c_1, c_2) - (\mu_1 - e z_1 \psi) c_1 - (\mu_2 + e z_2 \psi) c_2\right)  + \nonumber\\
  +& \int_V\!\!\!d^3{\bf r} \left( \frac{{\bf D}^2}{2 \varepsilon} -
     \psi \bnabla\cdot {\bf D} \right).
\end{align}
Invoking now the identity eq.~(\ref{dfgwy}), discarding the boundary
terms and minimizing with respect to $\bf D$, we get the final form of
the inhomogeneous thermodynamic potential
\begin{equation} {\cal F} [\psi] = - \!\!\!\int_V\!\!\!d^3{\bf r}
  \left( {\textstyle\frac12} {\varepsilon} (\bnabla\psi)^2 + p(\mu_1 - e
    z_1 \psi, \mu_2 + e z_2 \psi)\right).
  \label{gfiwq}
\end{equation}
In the case of charged boundaries one needs to add a surface term
$\oint_S \psi D_n dS$, where $D_n$ is the normal component of the
electric displacement field at the surface, to the above
equation. While the derivation of eq.~(\ref{gfiwq}) proceeded entirely
on the mean-field level, it can be extended to the case when the
Coulomb interactions are included exactly and the mean potential
becomes the fluctuating local potential in a functional integral
representation of the partition function \cite{Wiegel}.

Let us note that the signs of the electrostatic terms in
eq.~(\ref{gfiwq}) are consistent with the definition of the grand
canonical partition function, i.e. $\Omega = - p V$, with
$\Omega(\lambda, \beta) = \sum_{N=1}^\infty \lambda^N Q(N, \beta)/N!$,
where $Q(N, \beta)$ is the canonical partition function for $N$
particles.  The absolute activity is defined as
$\lambda = e^{\beta \mu}$. Since electrostatic interactions enter with
a Boltzmann factor,
$\lambda = e^{\beta \mu} \longrightarrow e^{\beta \mu \mp e z_{1,2}
  \psi}$, where $-$ is valid for positive and $+$ for negative ions.

For any equation of state $p(\mu_1, \mu_2)$ or indeed any model free
energy $f(c_1, c_2)$ of the reference uncharged system, one now needs
to evaluate the proper chemical potentials of the binary components
from eq.~(\ref{ekrw}), make a substitution
$$ \mu_{1,2} \longrightarrow \mu_{1,2} \mp e z_{1,2} \psi$$and
finally derive the Euler-Lagrange equation for the local electrostatic
potential of the form
\begin{equation} {\varepsilon} \bnabla^2 \psi - \frac{\partial p(\mu_1
    - e z_1 \psi, \mu_2 + e z_2 \psi)}{\partial \psi} = 0,
  \label{beiruwy}
\end{equation}
which generalizes a form derived within a symmetric lattice gas
approximation \cite{Trizac}.  Invoking furthermore the Gibbs-Duhem
relation
$$c_{1,2} =
\frac{\partial p}{\partial \mu_{1,2}}$$ we derive the Poisson equation as
\begin{align}
  \frac{\partial p(\mu_1 - e z_1 \psi, \mu_2 + e z_2
  \psi)}{\partial \psi} =& - e z_1  \frac{\partial p}{\partial \mu_{1}} + e z_2 \frac{\partial p}{\partial \mu_{2}} =\nonumber\\
                         & = - e (z_1c_1 - z_2c_2) =q. 
                           \label{befjkw}
\end{align}
where $q$ is the local charge density.  Note that the charge density is a derivative 
w.r.t. potential of a single function, a simple test of consistency for any proposed theory. 
Together with eq.~(\ref{beiruwy}) this constitutes a
{\em generalisation of the Poisson-Boltzmann theory} for any model of the fluid expressible via an
equation of state in the local thermodynamic approximation. This also generalizes some results previously
derived only for the lattice gas.

In the case of a single or two planar surfaces, with a normal in the
direction of the $z$-axis, so that $\psi({\bf r}) = \psi(z)$, the
Poisson-Boltzmann equation possesses a first integral of the form
\begin{equation} {\textstyle\frac12} {\varepsilon} \psi'^2(z) -
  p(\mu_1 - e z_1 \psi(z), \mu_2 + e z_2 \psi(z)) = -p_0
  \label{mhkl}
\end{equation}
where $p_0$ is an integration constant equal to the osmotic pressure of the ions and determined by the boundary
conditions.  The disjoining (interaction) pressure for two charged surfaces, $\Pi$, is then
obtained by subtracting the bulk contribution from the osmotic
pressure $p_0$. The first integral of the Euler-Lagrange equation can be
used to construct an explicit 1D solution, $\psi = \psi(z)$ by
quadrature.

In the limiting case of an ideal gas, with the van't Hoff equation of
state $p(c_1, c_2) = k_B T(c_1 + c_2)$ it is straightforward to see
that the above theory reduces exactly to the Poisson-Boltzmann
approximation \cite{COCIS}. Furthermore, for the binary, symmetric
lattice-gas
\begin{equation}
  p(c_1, c_2) = - \frac{k_B T}{a^3}~\log{(1 -  a^3 (c_1 + c_2))} \label{eq:korn}
\end{equation}
where $a$ is the cell size \cite{Danpaper}, the above formalism yields
the results discussed at length by Kornyshev
\cite{Kornyshev-rev}. From eq.~(\ref{eq:korn}) we also see one of the
weaknesses of the lattice gas approach, as the pressure diverges only
{\sl very weakly}\/ at close packing. We will compare with a more
realistic equation of state later in this paper.

\section{Asymmetric lattice gas}

We start with the free energy density of mixing for a three component
lattice gas system composed of species "1" at concentration $c_1$,
itself composed of $N_1$ subunits, and species "2" at concentration
$c_2$, itself composed of $N_2$ subunits, in a solvent of (water)
molecules of diameter $a$. It can be expressed rather
straightforwardly in terms of the volume fractions $\phi_1, \phi_2$
after realizing that it is equivalent to the problem of polydisperse
polymer mixtures on the Flory-Huggins lattice level
\cite{Teraoka}. For a two component system the free energy of mixing
can be derived simply as \cite{Muthu}
\begin{align}
  &\frac{f(\phi_1, \phi_2)~a^3}{k_B T} = \frac{\phi_1}{N_1} \log{\phi_1} + \frac{\phi_2}{N_2} \log{\phi_2} + \nonumber\\
  & + (1 - \phi_1-\phi_2)\log{(1 -\phi_1-\phi_2)},
\end{align}
where the volume fractions $\phi_1, \phi_2$ are defined as
\begin{equation}
  \phi_{1,2} = a^3 c_{1,2} N_{1,2} = R_{1,2}^3 c_{1,2},
\end{equation}
and $N_{1,2} = {(R_{1,2}/a)}^3$ measures the relative volumes of
species 1 and 2, with radii $R_{1,2}$, compared to the solvent with
radius $a$. While the size-symmetric lattice gas has a venerable
history (for an excellent review see Ref. \cite{Bazant}) there have
been fewer previous attempts to master the lattice gas mixtures in the
context of size-asymmetric electrolytes
\cite{Eigen,Lipfert,Zhou,Biesheuvel,Siber} and the simple connection
with the entropy of lattice polymers has apparently not been noted
before.

The chemical potential is then obtained as
\begin{equation}
  \mu_{1,2} = \frac{\partial f(c_1, c_2)}{\partial c_{1,2}} = \frac{\partial f(\phi_1, \phi_2)}{\partial \phi_{1,2}} a^3 N_{1,2},
\end{equation}
that can be evaluated explicitly yielding
\begin{equation}
  \mu_{1,2} = \log{\phi_{1,2}} + 1 - N_{1,2} \left(\log{(1 -\phi_1-\phi_2) + 1}\right).
  \label{efjkw}
\end{equation}
The Legendre transform eq.~(\ref{dapois}) then yields the osmotic
pressure, again as a function of both volume fractions
\begin{align}
  &- \frac{p(\phi_1, \phi_2)~a^3}{k_B T} = \log{(1 -\phi_1-\phi_2)} + \nonumber\\
  & + \phi_1\left( 1 - \frac{1}{N_1}\right) + \phi_2\left( 1 - \frac{1}{N_2}\right).
\end{align}
The form of this result is revealing as it states that the osmotic
pressure is basically the lattice gas pressure of a symmetric mixture,
corrected by the fact that $N_{1,2}$ subunits of the species "1" and
"2" do not represent separate degrees of freedom. Obviously, for a
symmetric system with $N_{1,2} = 1$ this reduces exactly to the
lattice gas symmetric binary mixture expression,
eq.~(\ref{eq:korn}). 

Introducing
$\tilde\mu_{1,2} = \mu_{1,2} + N_{1,2} -1$ we can rewrite
eq.~(\ref{efjkw}) as
\begin{equation}
  \phi_{1,2} = {(1 -\phi_1-\phi_2)}^{N_{1,2}}~e^{\tilde\mu_{1,2}}
  \label{eqrst}
\end{equation}
Using this relation we can derive an explicit equation for
\begin{equation}
u = (1 -\phi_1-\phi_2) 
\label{dglv}
\end{equation}
of the form
\begin{equation}
  u \left(1 + u^{N_1-1}e^{\tilde\mu_{1}} +
    u^{N_2-1}e^{\tilde\mu_{2}}\right) = 1, \label{eq:bhdc}
\end{equation}
that yields $u = u(\tilde\mu_{1},\tilde\mu_{2}; N_1, N_2)$.  This
allows us to finally write the osmotic pressure as a function of the
two densities
\begin{align}
  &- \frac{p(c_1, c_2)~a^3}{k_B T} = \log{(1 -a^3 (c_{1} N_{1} + c_{2}
    N_{2})) } + \nonumber\\
  & + a^3 c_{1} \left( N_1 - 1\right) + a^3 c_{2} \left( N_2 - 1\right),
\end{align}
or of the two chemical potentials through
$u = u(\tilde\mu_{1},\tilde\mu_{2}; N_1, N_2)$ as
\begin{align}
  &- \frac{p(\mu_1, \mu_2)~a^3}{k_B T} = \log{u} + \nonumber\\
  & + u^{N_1} e^{\tilde\mu_{1}}\left(1 - \frac{1}{N_1}\right) + u^{N_2} e^{\tilde\mu_{2}} \left(1 - \frac{1}{N_2}\right).
\end{align}
In the case of ions of the same size, we can set without any loss of
generality, that $N_1 = N_2 = 1$, so that
\begin{align}
  - \frac{p(\mu_1, \mu_2)~a^3}{k_B T} = \log{u(\tilde\mu_{1},\tilde\mu_{2})}  =
  - \log{\left(1+e^{\tilde\mu_{1}}+e^{\tilde\mu_{2}}\right)},\nonumber\\
\end{align}
a standard expression for the symmetric lattice gas \cite{Orland}.

Above equations present a complete set of relations satisfied by the
asymmetric lattice gas, being a mixture of two differently sized
ions. The addition of mean-field electrostatic interactions
eq~(\ref{beiruwy}) then modifies solely the chemical potentials so
that
\begin{equation}
  p(\mu_1, \mu_2) \longrightarrow  p(\mu_1 - e z_1 \psi, \mu_2 + e z_2 \psi)
\end{equation}
if the two species are oppositely charged, which we assume. The
corresponding Poisson-Boltzmann equation is then obtained from
eq.~\ref{beiruwy} and eq.~\ref{befjkw} in the form
\begin{eqnarray}
  {\varepsilon} \bnabla^2 \psi\!\!\!&=&\!\!\!- e\left( z_1 {\partial_{\mu_1}} - z_2 {\partial_{\mu_2}}\right)  p(\mu_1 - e z_1 \psi, \mu_2 + e z_2 \psi) =\nonumber\\
                                    &=& - e
                                        (\frac{z_1}{N_1}\phi_1(\psi) -
                                        \frac{z_2}{N_2}\phi_2(\psi)).
                                        \label{befjkw1}
                                        \label{aeds}
\end{eqnarray}
where $\phi_{1,2}(\psi)$ are obtained from eq.~\ref{eqrst} and
\ref{eq:bhdc} with
$\tilde\mu_1 \longrightarrow \tilde\mu_1- e z_1 \psi, \tilde\mu_2
\longrightarrow \tilde\mu_2+ e z_2 \psi$.

In complete analogy with the case of  polyelectrolytes with added salt \cite{Ramanathan}  
it is clear that electroneutrality of the asymmetric 
lattice gas in the bulk is achieved only if it is held at a non-zero electrostatic potential, $\psi_0$, 
that can be obtained from eq.~(\ref{aeds}) in an implicit form 
\begin{equation} 
(N_1-N_2) \log{u(\psi_0)} = - ({\tilde\mu_1-\tilde\mu_2}) + \log{\frac{N_1 z_2}{N_2 z_1}}.
    \label{78690}
\end{equation}
In what follows we then simply displace the origin of the electrostatic potential by $\psi_0$, the Donnan potential, interpreted as the change in electrostatic potential across the bulk reservoir - ionic liquid interface, or equivalently as a Lagrange multiplier for the constraint of global electroneutrality \cite{Denton}.

\section{Asymptotic behaviour of the lattice gas model}

We now consider the forms of the general equations derived above in the limiting cases of small and large electrostatic potential of the lattice gas model:

\subsection{Small potential and screening length}

In the limit of $\psi \rightarrow 0$, one can derive
\begin{align}
  &p(\mu_1 + e z_1 \psi, \mu_2 - e z_2 \psi) = p(\mu_1, \mu_2) ~+ \nonumber\\
& \qquad - e\left( z_1 {\partial_{\mu_1}} - z_2 {\partial_{\mu_2}}\right)p(\mu_1, \mu_2) \psi ~+ \nonumber\\
  & \qquad + {\textstyle\frac12} e^2 {\left( z_1 {\partial_{\mu_1}} - z_2 {\partial_{\mu_2}}\right)^2} p(\mu_1, \mu_2)~\psi^2 + {\cal O}(\psi^3)
    \label{erqtw}
\end{align}
where we took into account eq.~(\ref{befjkw}). Just as in the full non-linear case, see above, the term linear in $\psi$ 
is connected with the displaced electrostatic potential, eq.~(\ref{78690}). 
The linearized form of $\psi_0$  is then obtained approximately as
\begin{equation}
\psi_0 = \frac{e\left( z_1 {\partial_{\mu_1}} - z_2 {\partial_{\mu_2}}\right)p(\mu_1, \mu_2)}{e^2 {\left( z_1 {\partial_{\mu_1}} - z_2 {\partial_{\mu_2}}\right)^2} p(\mu_1, \mu_2)}.
\end{equation}
Obviously the expansion of the pressure for small values of the electrostatic potential is then quadratic in the difference $\psi - \psi_0$.

Furthermore, the Hessian of the pressure $p(\mu_1, \mu_2)$ is positive definite, i.e.
\begin{equation}
  \frac{e^2}{\varepsilon} \left( z_1^2\frac{\partial^2 p}{\partial \mu_{1}^2} - 2 z_1z_2\frac{\partial^2 p}{\partial \mu_{1}\mu_{2}} + z_2^2\frac{\partial^2 p}{\partial \mu_{2}^2}\right) = \kappa^2 > 0,
\end{equation}
while from eq.~(\ref{gfiwq}) it follows that $\kappa$ is nothing but the
{\em inverse Debye length} expressed through the second derivatives of the
pressure with respect to the chemical potentials of both charged
species. Since the curvature tensor of the Legendre transform is the
inverse of the curvature tensor of the function itself \cite{Zia}, we
can write
\begin{equation}
  \sum_m \frac{\partial^2 p(\mu_1, \mu_2)}{\partial \mu_{i} \partial \mu_{m}} \frac{\partial^2 f(c_1, c_2)}{\partial c_{m} \partial c_{k}} = \delta_{ik}
\end{equation}
where all the matrices are $2 \times 2$. From here it follows rather
straightforwardly that
\begin{align}
  \kappa^2 = &\frac{e^2}{\varepsilon} \frac{z_2^2\frac{\partial^2 f}{\partial c_{1}^2} + 2 z_1z_2\frac{\partial^2 f}{\partial c_{1}\partial c_{1}}+ z_1^2\frac{\partial^2 f}{\partial c_{2}^2}}{\frac{\partial^2 f}{\partial c_{1}^2} \frac{\partial^2 f}{\partial c_{2}^2} - {\left(\frac{\partial^2 f}{\partial c_{1}\partial c_{2}}\right)}^2}  \nonumber\\
  =& 4\pi \ell_B N_1 N_2 \frac{u (z_1^2c_1 + z_2^2c_2) + a^3 c_1c_2
     {({z_1}{N_2} + {z_2}{N_1})}^2 }{(1 + (N_1-1)a^3 N_1 c_1 + (N_2-1)a^3 N_2 c_2)},\nonumber\\
\end{align}
where we introduced the Bjerrum length
$\ell_B = e^2/(4\pi \varepsilon k_B T)$.  In general the Debye length
is therefore {\em not} a linear function of the concentrations. For
the symmetric lattice gas, $N_1 = N_2 = 1$, and taking into account
the definition eq. \ref{dglv}, the above result reduces to
$\kappa^2 = 4\pi \ell_B \left( z_1^2c_1 + z_2^2c_2 - a^3{(z_1c_1 -
    z_2c_2)}^2\right)$,
which for bulk electroneutrality reduces further to the standard Debye
expression\cite{Orland}.


\subsection{Large potential and close packing}

The limits for $\psi \rightarrow \pm\infty$ of a lattice gas can be
derived as
\begin{equation}
  u = \left\{
    \begin{array}{ll}
      e^{-(\tilde\mu_{1}+ e z_1\psi)/N_1} & \psi \rightarrow +\infty   \\
      e^{-(\tilde\mu_{2}- e z_2\psi)/N_2} & \psi \rightarrow -\infty 
    \end{array}
  \right. 
\end{equation}
implying
\begin{equation}
  \phi_{1,2}(\psi \rightarrow \infty) =  \left\{
    \begin{array}{l}
      1   \\
      e^{-(\tilde\mu_{1} + e z_1\psi) N_2/N_1 + (\tilde\mu_{2} - e z_2\psi)}
    \end{array}
  \right..
\end{equation}
and
\begin{equation}
  \phi_{1,2}(\psi \rightarrow -\infty) =  \left\{
    \begin{array}{l}
      e^{-(\tilde\mu_{2} - e z_2\psi) N_1/N_2 + (\tilde\mu_{1} - e z_1\psi)}  \\
      1
    \end{array}
  \right..
\end{equation}
where the upper formula is for "1" and the lower for "2". Thus
\begin{align}
  &- \frac{p(\psi \rightarrow \infty)~a^3}{k_B T} = -(\tilde\mu_{1} + e z_1\psi)/N_1 + \nonumber\\
  & + \left( 1 - \frac{1}{N_1}\right) + e^{-(\tilde\mu_{1}+ e z_1\psi) N_2/N_1} e^{(\tilde\mu_{2}- e z_2\psi)}
    \left( 1 - \frac{1}{N_2}\right)\nonumber\\
\end{align}
and
\begin{align}
  &- \frac{p(\psi \rightarrow -\infty)~a^3}{k_B T} = -(\tilde\mu_{2}- e z_2\psi)/N_2 + \nonumber\\
  & + e^{-(\tilde\mu_{2}- e z_2\psi) N_1/N_2} e^{(\tilde\mu_{1}+ e z_1\psi) } \left( 1 - \frac{1}{N_1}\right) + 
    \left( 1 - \frac{1}{N_2}\right)\nonumber\\
\end{align}

The most striking feature of these limits is the linear behaviour of
$p(\psi)$ for large positive or negative potentials, which gives rise
to a V-like curve for symmetric particles. This linear behaviour is
linked to the saturation of close packing of the lattice particles against a high
potential surface. For particle of unequal size the two branches of $p(\psi)$ have different slopes.

\section{A dense two component lattice gas}
For some cases in the theory of ionic liquids one can assume dense
packing, without any intervening solvent, so that
$\phi_1 + \phi_2 = 1$. The corresponding free energy can then be cast
into a simplified form
\begin{eqnarray}
  &&\frac{f(\phi_1, \phi_2 = 1 - \phi_1)~N_2 a^3}{k_B T} = \frac{\phi_1}{M} \log{\phi_1} + {\phi_2} \log{\phi_2}.   \nonumber\\
\end{eqnarray}
where $M = N_1/N_2$ is the effective size of the species "1" compared
to species "2". This implies furthermore that
\begin{equation}
  \frac{\partial f}{\partial \phi_1} + \frac{\partial f}{\partial \phi_2} = 0 \longrightarrow \mu_1 + M \mu_2 = 0. 
\end{equation}
The equation analogous to eq.~(\ref{eq:bhdc}) then assumes the form
\begin{equation}
  \phi_1 = {(1 - \phi_1)}^{M} e^{\mu_1 - 1 + M}
\end{equation}
and the Legendre transform of the free energy density follows as
\begin{eqnarray}
  \!\!\!&&\!\!\!- \frac{p(\phi_1, \phi_2= 1 - \phi_1)~N_2 a^3}{k_B T} = \log{\phi_2} + \phi_1\left( 1 - \frac{1}{M}\right). \nonumber\\
\end{eqnarray}
The Poisson-Boltzmann equation is then cast into a simplified form
\begin{align}
  \varepsilon\nabla^2\psi =& -e\left( z_1 {\partial_{\mu_1}} - z_2 {\partial_{\mu_2}}\right) p(\mu_1 - ez_1\psi, \mu_2 + ez_2\psi) = \nonumber\\
                           & = \frac{1}{N_1} \left( (z_1 + M z_2) \phi_1 - M z_2\right),
\end{align}
and the charge density is constrained to be between
$- \frac{z_2}{N_2}$ and $ \frac{z_1}{N_1} $.


\section{Asymmetric Carnahan-Starling approximation}

In order to show the interest and generality of our local thermodynamic approach we will
now apply it in the case of the Carnahan - Sterling approximation for
asymmetric binary hard sphere mixtures \cite{Mansoori}. For a bulk, uncharged hard sphere
fluid the \CS~  approximation is "almost exact".

The excess pressure in the \CS~ approximation derived via the "virial equation" is then equal
to
\begin{equation}
  \frac{p_{exc}(c_1, c_2)}{c k_B T}  = \frac{(1+\xi + \xi^2) - 3\xi (y_1+\xi y_2) - 3 \xi^3 y_3}{{(1 - \xi)}^3}
\end{equation}
where $c_{1,2}$ are the densities of the two components and
\begin{equation}
  \xi_{1,2} = \frac{4\pi}{3} R_{1,2}^3 c_{1,2} \qquad {\rm and} \qquad \xi = \xi_1 + \xi_2,
\end{equation}
where $R_{1,2}$ are the hard sphere radii of the two
species. Furthermore
\begin{align}
  y_1 =& \Delta_{12} \frac{R_1 + R_2}{\sqrt{R_1 R_2}} \qquad 
         y_2 = \Delta_{12} \frac{\xi_1 R_2  + \xi_2  R_1 }{\sqrt{R_1 R_2}\xi} \nonumber\\
  y_3 =& {\left( {\left(\frac{\xi_1}{\xi}\right)}^{2/3} +  {\left(\frac{\xi_2}{\xi}\right)}^{2/3}\right)}^3
\end{align}
with
\begin{equation}
  \Delta_{12} = \frac{\sqrt{\xi_1 \xi_2}}{\xi} \frac{{(R_1 - R_2)}^2}{R_1 R_2}.
\end{equation}
The excess free energy then follows as
\begin{align}
  \frac{f_{exc}(c_1, c_2)}{c k_B T} =& -{\textstyle\frac32} \left( 1 - y_1 + y_2 + y_3\right) + \nonumber\\
                                     & + \frac{3 y_2 + 2 y_3}{1 - \xi} + {\textstyle\frac32}\frac{1 - y_1 - y_2 - {\textstyle\frac13}  y_3}{{(1 - \xi)}^2} + \nonumber\\
                                     & + (y_3 - 1) \log{(1 - \xi)}, \label{eq:denom}
\end{align}
It is now straightforward to obtain the chemical potentials from the
free energy as $\mu_{1,2} = \mu_{1,2}(c_1, c_2)$, invert them and then
obtain the pressure equation as $p = p(\mu_1, \mu_2)$.

\begin{figure}
  \includegraphics[scale=0.45]{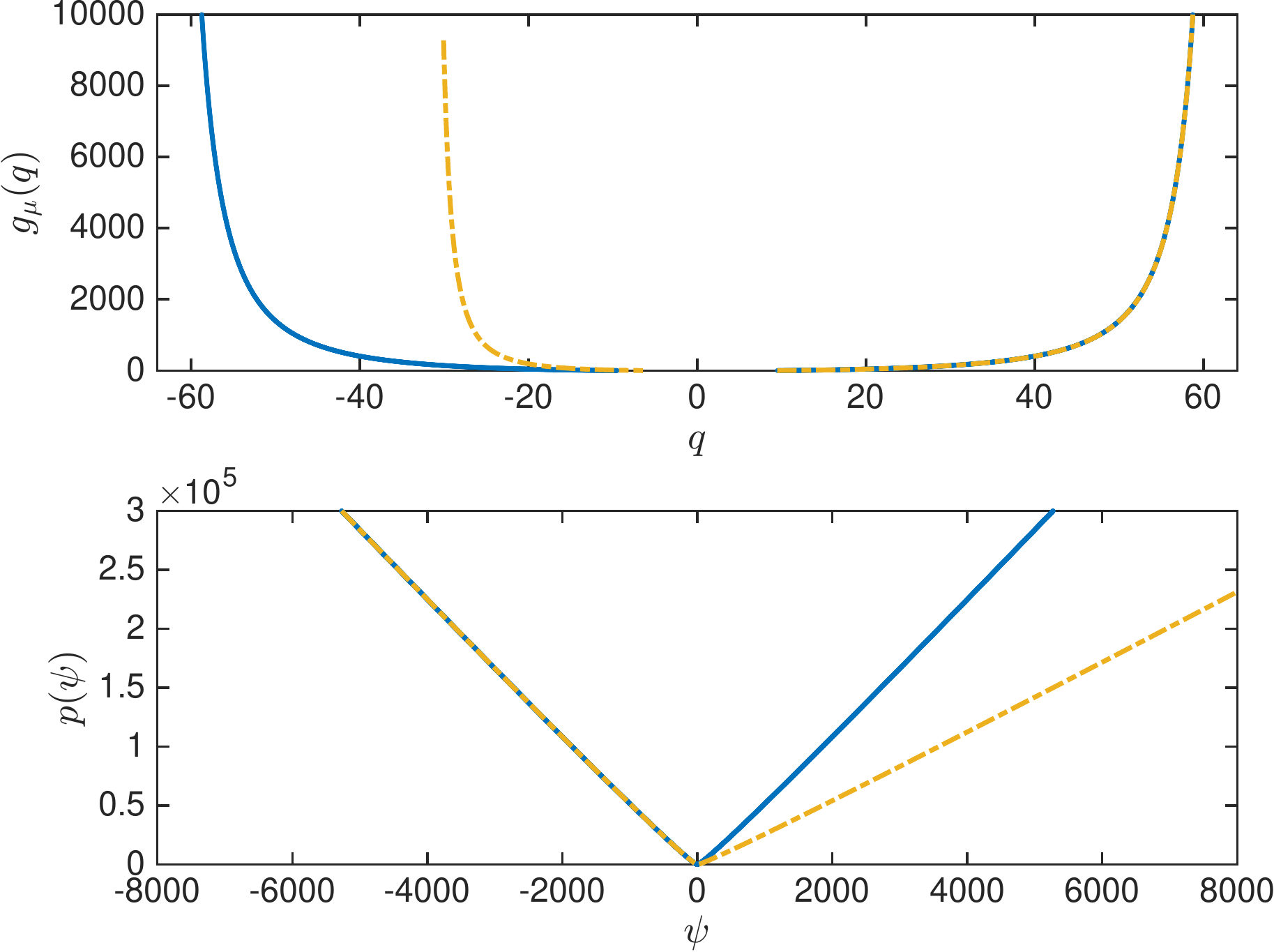}
  \caption{\label{fig:CS} Top: the function $g_\mu(q)$ for a {\CS}
    fluid. It is an effective potential at imposed local charge
    density. The divergences correspond to the  close packing of the
    fluid. Blue: symmetric particle volumes. Gold: asymmetric fluid with
    larger negative particles. This function is important in dual
    convex formulations of the Poisson-Boltzmann equation, see
    eq.~(\ref{eq:dual}). Bottom: the function $p(\psi)$ for the same
    two sets of fluid parameters.  Blue: Seen from afar the function
    displays a characteristic V-like behaviour for symmetric
    particles. The larger particles, gold, give a smaller slope in the
    pressure function. The central part of the figure for symmetric
    particles is examined more closely in figure~\ref{fig:zoom} }
\end{figure}



\begin{figure}
  \includegraphics[scale=0.47]{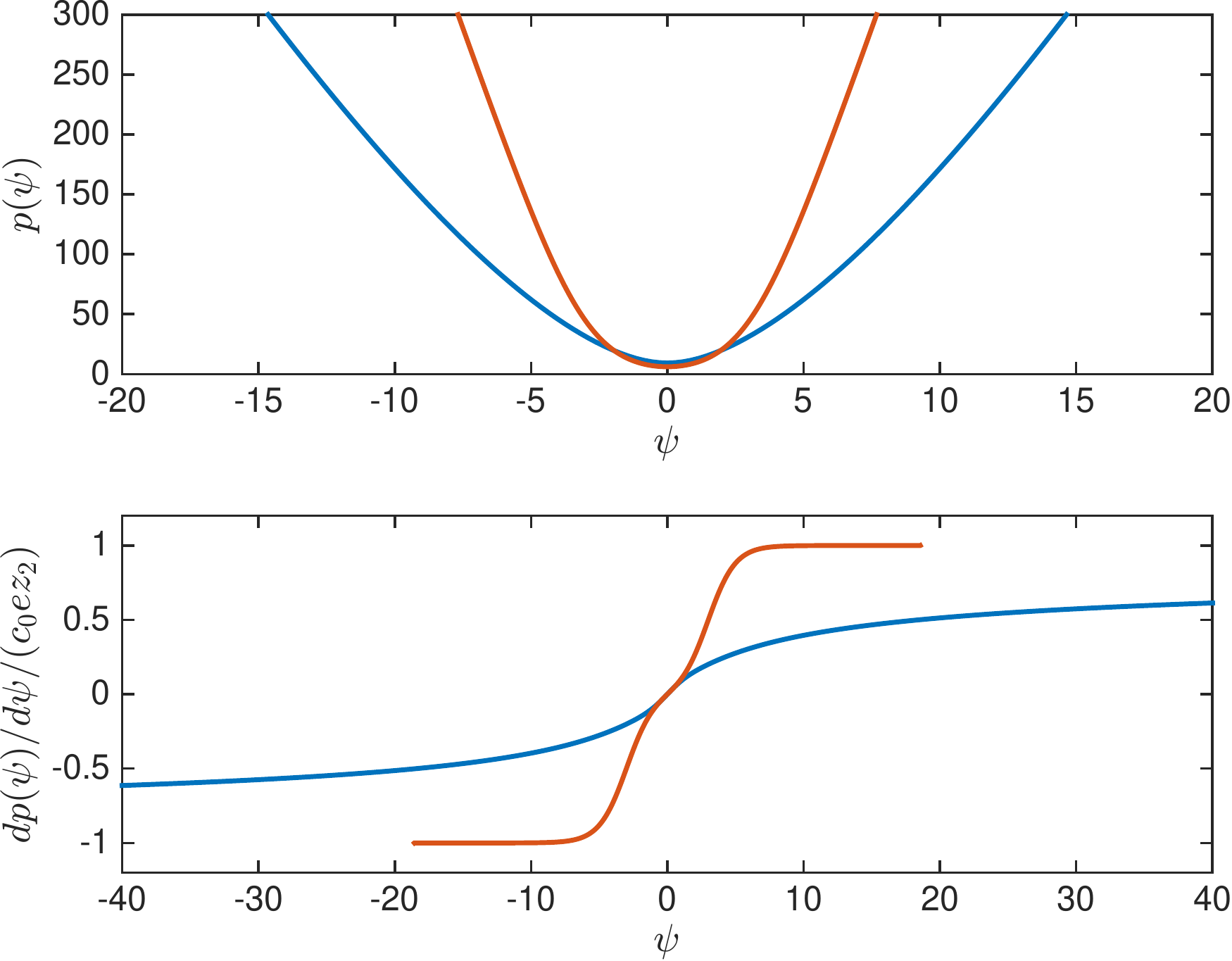}
  \caption{\label{fig:zoom} Comparison of symmetric {\CS} and lattice
    gas forms near $\psi=0$. Top $p(\psi)$, bottom charge density
    $-q(\psi)= dp/d\psi$.  {\CS} in blue, converges very slowly to the
    slope final value, (see figure.~\ref{fig:extrapol}), and is much
    broader at minimum compared to the lattice gas curve
    (red). Physical parameters are such that the blue and red curves
    have asymptotically the close packing charge density; the chemical
    potentials have been tuned to give the same Debye length in the
    bulk (corresponding to the same curvature at the origin in the
    curve for $p$). Despite this double matching the curves are very
    different in functional form. The curve $q(\psi)$ for the {\CS}
    fluid converges extraordinarily slowly compared with the lattice
    gas approximation.}
\end{figure}


\begin{figure}
  \includegraphics[scale=0.47]{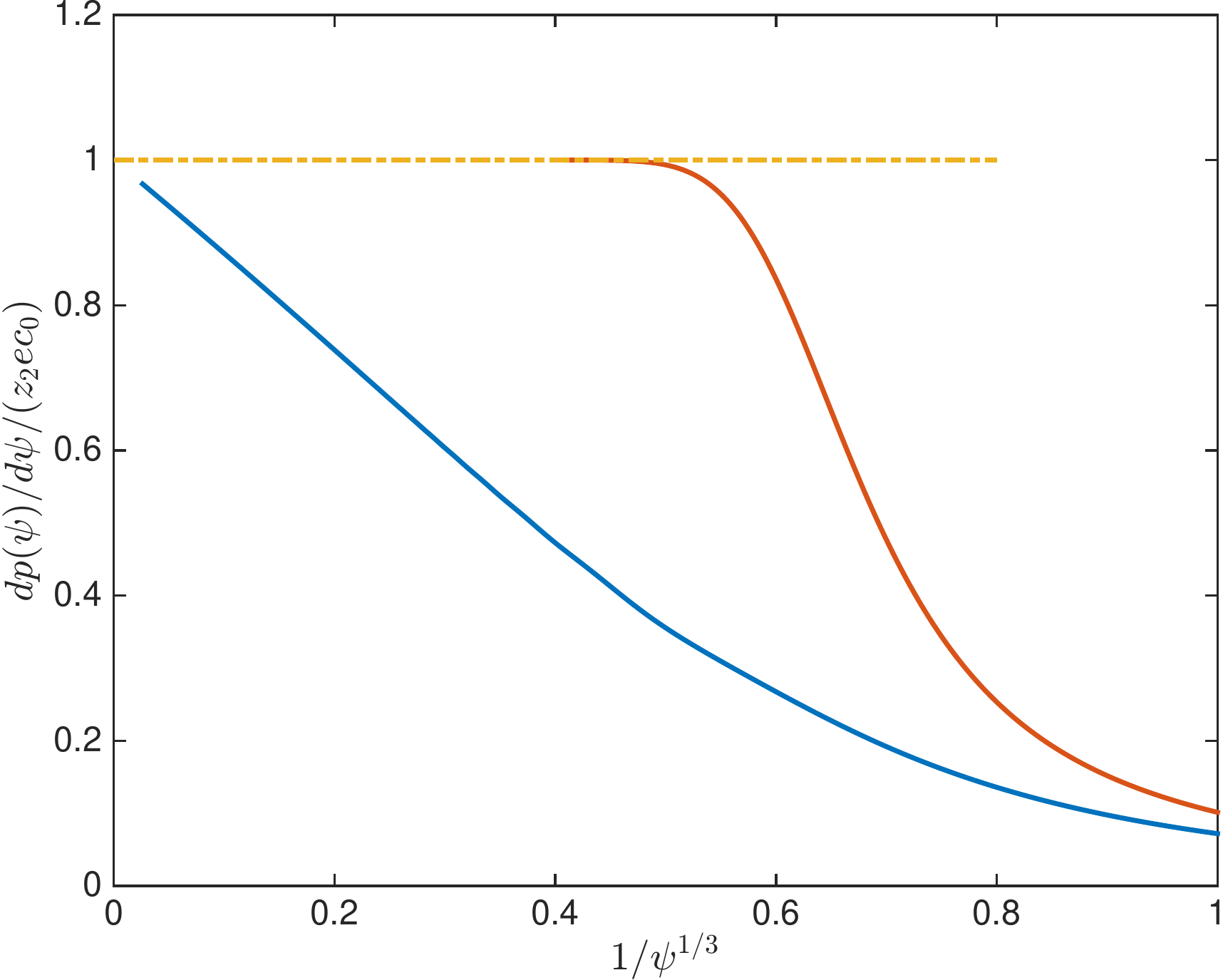}
  \caption{\label{fig:extrapol} Study of the evolution of $-q=dp(\psi)/d\psi$ in variables adapted
    to the {\CS} fluid, (blue).  The large field behaviour is linear in $1/\psi^{1/3}$, demonstrating the
    correctness of arguments leading to eq.~(\ref{eq:csLeg2}).  Same data as fig.~\ref{fig:zoom} for
    large positive potentials. The lattice gas model (red) gives much more rapid cross-over to
    saturation at large potentials. Horizontal line is to guide the eye
    and corresponds to the close packing charge density.}
\end{figure}


\section{Asymptotic behaviour for the {\CS} free energy density}

In the limit of large potentials $\psi$, as occurs near an electrode,
the second, wrongly charged, component of the fluid is excluded and
the dominant physics is the packing of a single component system under
the constraints coming from the electrostatic interactions. In this
limit of $\xi \longrightarrow 1$ we can substitute $y_{1,2}=0$ and
$y_3=1$.  The most important divergence in this limit thus stems from
the denominator in eq.~(\ref{eq:denom}).

The free energy density of a \CS ~liquid near close packing has a
singularity of the form
\begin{equation} f(c) = \frac{c_0 k_B T }{{(1- c/c_0)}^2}
\end{equation}
$c_0$ is the close packing volume fraction of the component dominating
near the electrode.  With this assumption we can take the Legendre
transform of the most singular, diverging part of the free energy to
find the large potential limit of $p(\psi)$. For large positive $\psi$
(assuming that $e z_2 \psi \gg \mu$) this limit turns out to be
\begin{equation}
  p(\psi) = z_2e c_0 \psi - \frac{3}{2} {(2 c_0 k_B T)}^{1/3} {(z_2 e c_0
    \psi)}^{2/3} \label{eq:csLeg}
\end{equation}
where we have used the fact that negative ions of valence $z_2$
dominate.

In the high packing limit $p(\psi)$ is therefore linear in the
potential, and is given by the spatial charge density at close packing
exactly like for the lattice gas. However unlike the lattice gas the
approach to the high field limit is very slow
\begin{equation}
  -q(\psi) = z_2e c_0  -  \Big {( \frac{2 c_0 k_B T }{\psi} \Big)}^{1/3} {(z_2e c_0)}^{2/3}
  \label{eq:csLeg2}
\end{equation}
The spatial charge density is negative for large positive potentials.
The correctness of this law is demonstrated in
figure~\ref{fig:extrapol} which plots $(1/c_0 z_2 e) dp/d\psi$ as a
function of $\psi^{-1/3}$. The curve linearly extrapolates to
unity for large $\psi$. There is a very clear contrast with the case
of the lattice gas model where the cross-over to close packing occurs
for much smaller values of the potential.

\subsection{Solution for the high field {\CS} limit}

The solution for the generalized Poisson-Boltzmann equation in the
high field limit can be found from the solution of the integral
problem
\begin{equation}
  \int\!\!{d \psi}{\left( e c_0 z_2 \psi - \beta  {(e c_0 z_2 \psi)} ^{2/3} +p_0\right)}^{-1/2}= \sqrt{\frac{2}{\varepsilon}}\int \!\!dz
\end{equation}
with $\beta = 3{(2 c_0 k_B T )}^{1/3}/2$; we neglect $\mu$ compared to
$z_2e \psi$.  This integral can be transformed by substituting
${(e c_0 z_2 \psi)}^{1/3} =y$, giving

\begin{equation}
  3\int \!\frac{dy ~y^2}{ {(y^3 - \beta y^2 +p_0)}^{1/2} } =
  \sqrt{\frac{2e^2 c_0^2 z_2^2}{\varepsilon}}  \int\! dz,
\end{equation}
a form which can be solved by using elliptic functions. If we make the
further approximation that $p_0$ is small we can find much simpler
expressions:
\begin{align}
  z(\psi) - z_0 
  = & \frac{ \sqrt{2 \epsilon}}{ ec_0 z_2}   [  {(  {(z_2 e c_0 \psi)}^{1/3}-\beta)}^{3/2} +
      \nonumber \\
  3&\beta {( {(z_2 e  c_0 \psi)}^{1/3}-\beta)}^{1/2}   ] \label{eq:csfit}
\end{align}

\begin{figure}
  \includegraphics[scale=0.45]{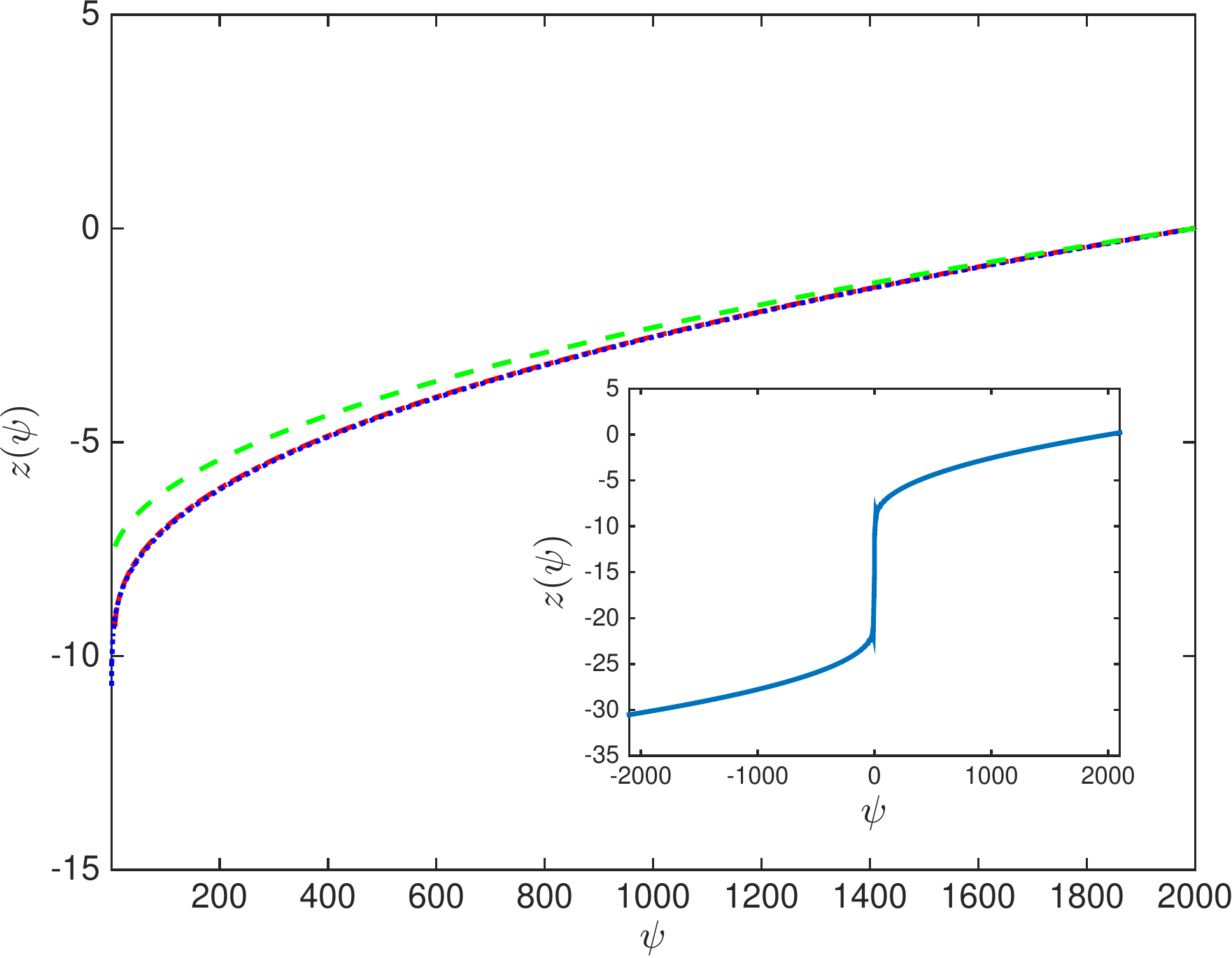}
  \caption{\label{fig:csfit} {\CS} fluid. Blue: exact numerical
    calculation of the function $z(\psi)$. The full curve is given in
    the inset. The high field limit studied in the main panel. Red:
    approximation valid for large fields from
    eq.~(\ref{eq:csfit}). Green profile calculated from assuming
    perfect packing of the fluid against electrode.  Blue and red
    curves almost perfectly correspond, with no extra fitting
    parameters. } \label{fig:exact}
\end{figure}

Here, $z(\psi)$ gives the distance from a plate which corresponds to a
potential $\psi$. It is obviously the inverse function of
$\psi(z)$. We perform a ``numerically exact'' calculation of the curve
$\psi(z)$ in fig.~(\ref{fig:exact}) in the inset, where we place a
positve electrode at $z=0$. The main figure of fig.~(\ref{fig:exact})
contains three curves: The blue curve explodes part of the inset and
is overlayed with a red curve corresponding to
eq.~(\ref{eq:csfit}). On this scale the results are
indistinguishable. The green curve is  evaluated by assuming
perfect packing of the fluid against the
electrode. Eq.~(\ref{eq:csfit}) is clearly a much better description
of the high electrostatic potential physics.

Eq.~(\ref{eq:csfit}) can also be combined with eq.~(\ref{eq:csLeg2})
to find $z(q)$ and thus the evolution of the spatial charge density
with distance from an electrode as well as the variation of the local
charge density with the potential.

\section{Differential Capacitance}

\begin{figure} 
\begin{minipage}[c][11cm][t]{.49\textwidth} 
  \vspace*{\fill}
  \centering
  \includegraphics[width=0.9\textwidth]{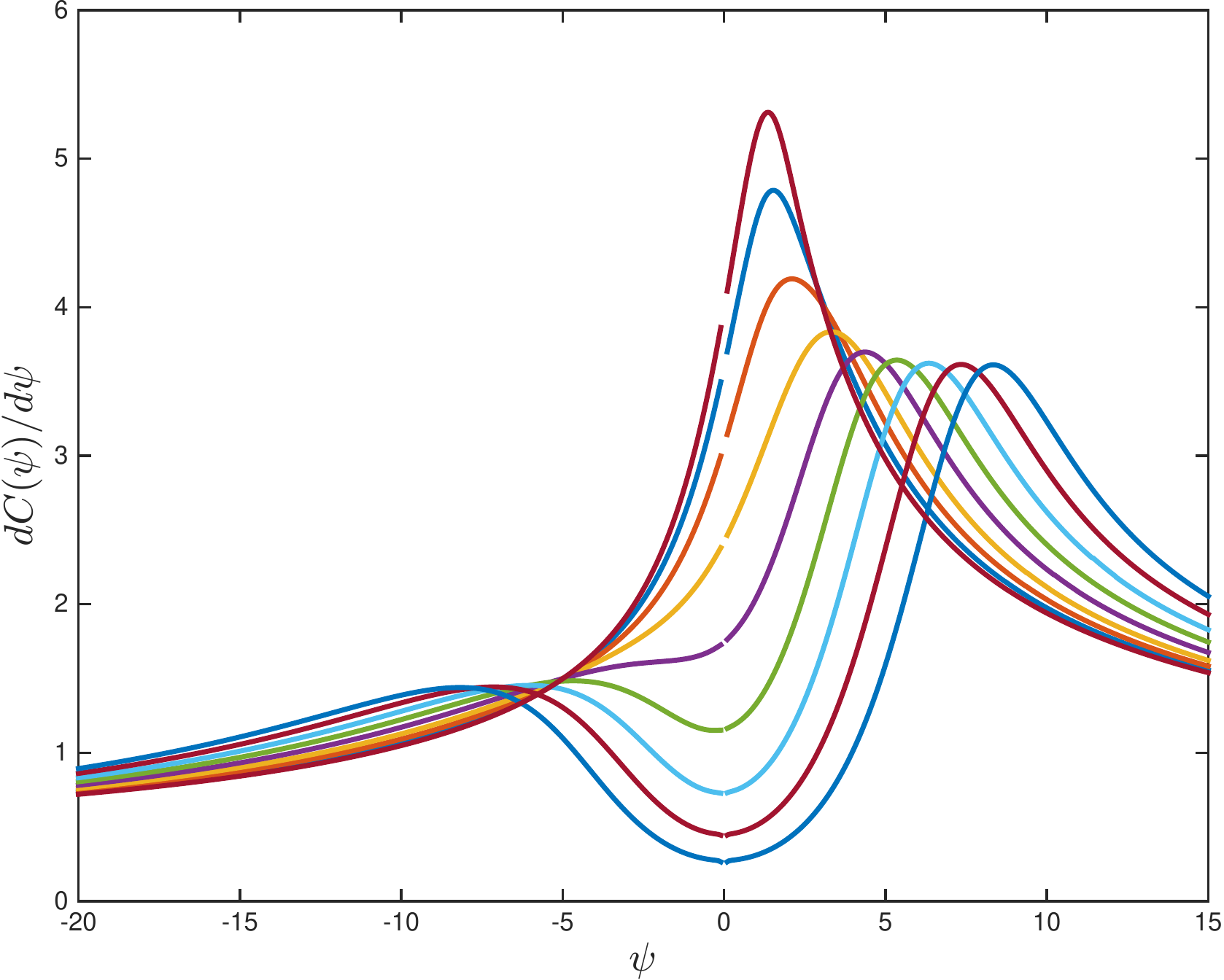}\par\vfill
  \includegraphics[width=0.9\textwidth]{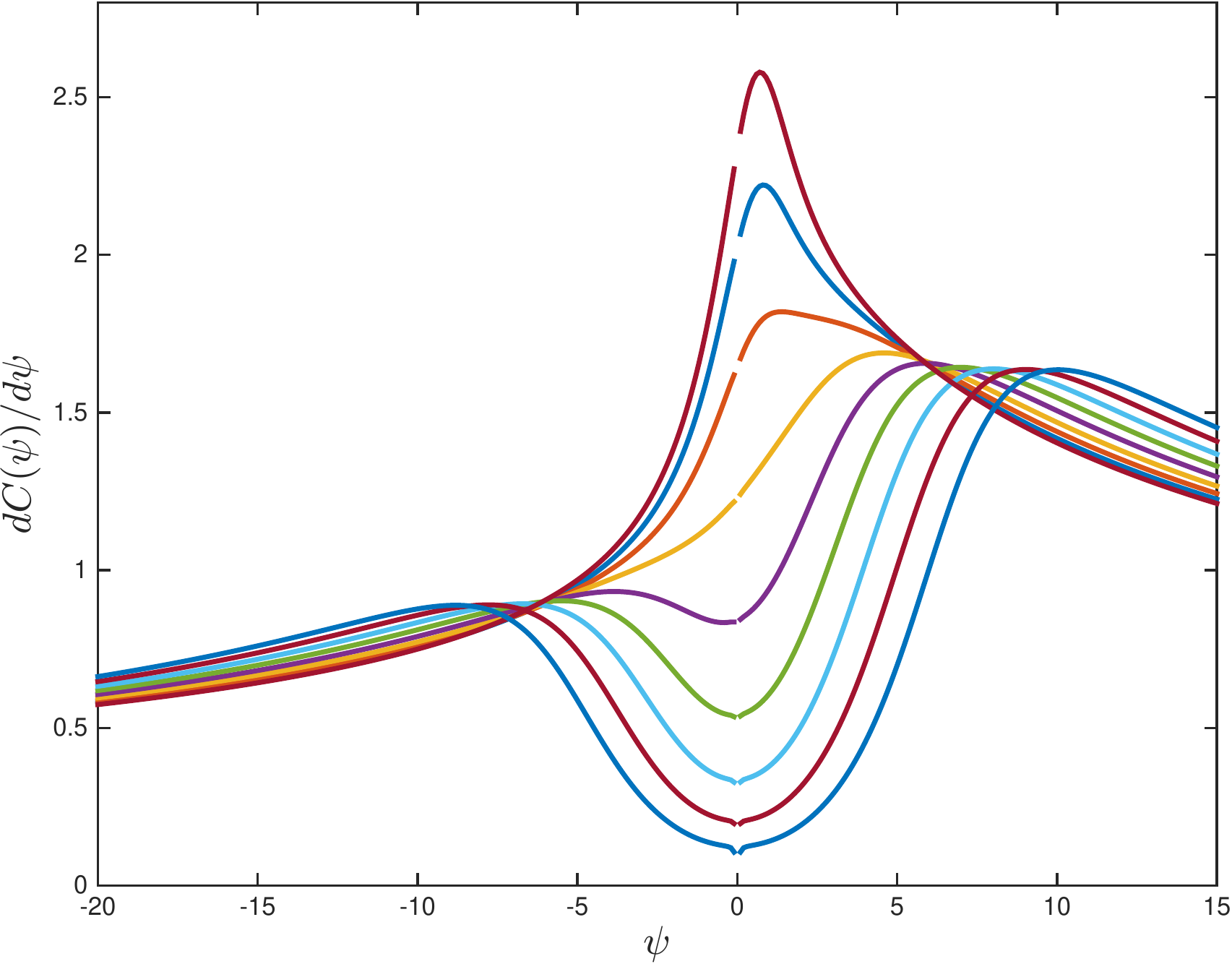}
\end{minipage}
\vskip 1.5cm
\caption{Top: calculation of the differential capacitance for a volume
  asymmetric lattice fluid. Bottom: volume asymmetric {\CS} fluid. From 
  eq.~(\ref{eq:grahame}). Each curve is for a different value of the
  chemical potential, $\mu$. Strongly negative chemical potentials
  give a minimum in the curves near $\psi=0$, together with two
  asymmetric maxima. As the chemical potential increases the curves
  develop a peak near (but not at) $\psi=0$. The chemical potential
  has been shifted such that the minimum of $p(\psi)$ is at
  $\psi=0$. In both cases the positively charged component has a
  volume $1.5^3$ larger than the negative component.}    \label{fig:cap}
\end{figure}

Together with the boundary condition $D_n = \sigma$, where $\sigma$ is
the surface charge density one can derive the equivalent of the
Grahame equation in the form
\begin{equation}
  \frac{\sigma^2}{{2 \varepsilon} } - p(\mu_1 - e z_1 \psi_0, \mu_2
  + e z_2 \psi_0) = -p_0 ,
\end{equation}
assuming that the bounding surface is located at $z=0$, i.e.
$\psi_0 = \psi(z=0)$. From the Grahame equation one can next derive
the differential capacitance $\cal C$ as
\begin{align}
  {\cal C}(\psi_0)  =  \frac{\partial \sigma(\psi_0)}{\partial \psi_0} = & \frac{\varepsilon e}{\sigma(\psi_0)} \left( - z_1\frac{\partial p}{\partial \mu_{1}} + z_2 \frac{\partial p}{\partial \mu_{2}}\right) = \nonumber\\
  =& \frac{- \varepsilon e (z_1 c_1 - z_ 2c_2)}{\pm\sqrt{2 \varepsilon
     (p-p_0) }},
\end{align}
with $\pm$ depending on the sign of the surface charge. Taking into
account the definition of the Bjerrum length, $\ell_B$
\begin{equation} {\cal C}^2(\psi_0) = 2\pi {k_B T} \ell_B
  \frac{(z_1c_1(\psi_0) - z_2c_2(\psi_0))^2}{{p(\psi_0) -p_0 }}.
\end{equation}
Invoking the Poisson-Boltzmann equation for this case, an alternative
form of the differential capacitance is
\begin{align}
  {\cal C}(\psi_0) =&  {\varepsilon}  \left( \log{\psi_0'}\right)' = \nonumber\\
  = &\sqrt{{2}{\varepsilon}} \frac{\partial}{\partial \psi_0}
      \sqrt{p(\mu_1 - e z_1 \psi_0, \mu_2 + e z_2 \psi_0)  -p_0 },
      \label{eq:grahame}
\end{align}
the form that we use in our numerical work. It is interesting to note
that even if we shift the minimum of the curve $p(\psi)$ to occur at
$\psi=0$ this does not imply that $\psi=0$ is also a stationary value
of the differential capacitance. This is clearly visible in the curves
of fig.~(\ref{fig:cap}) where in denser fluids the maximum of the
curves is shifted to positive potentials. We mark the position of the
minimum in $p(\psi)$ by a slight break in the solid lines. {This
  displacement of the maximum of the capacitance from the minimum of
  $p$ is trivially understood if one assumes that the expansion of
  $p(\psi)$ includes a term in $\psi^3$.}  {We see that the
  qualitative behaviour of the curves generated for the lattice model,
  as well as the {\CS} fluid are rather similar.}

\section{Conclusions}

By using general arguments based upon local thermodynamics, we generalized the Poisson-Boltzmann
mean-field theory of Coulomb fluids to the case where the reference, uncharged fluid need not be
ideal. We formulated the general theory in the particular cases of an asymmetric lattice gas and an
asymmetric {\CS} liquid that describe steric effects at various levels of approximations and are
particularly relevant for analysis of dense electric double layers as arise in the context of ionic
liquids or dense Coulomb fluids. Use of properties of Legendre transforms allows us to efficiently
translate between forms of the free energy; this includes a standard formulation in terms of
the electrostatic potential, and a dual formulation (see appendix) in terms of the electric displacement field.

We analyzed in detail the size and charge asymmetry and their
respective effects on the salient properties of electric double
layers.  As part of our analysis we also formulated an exact
thermodynamic description of an asymmetric lattice gas, derived within
the Flory-Huggins lattice approximation. This allows the lattice gas
approximation, which in its symmetric form already serves as the most
popular description of the steric effects in the context of the
Poisson-Boltzmann theory \cite{Bazant}, to be further extended to the
case of ubiquitous size-asymmetric dense ionic mixtures. It is
probably in this latter case that it will prove to be most useful
specifically in the context of ionic liquids \cite{Kornyshev-Fedorov}.

For the {\CS} fluid we have found an asymptotic form that gives a
rather simple analytic relation between potential and distance,
eq.~(\ref{eq:csfit}), as well as the relation between potential and
local charge density eq.~(\ref{eq:csLeg2}). It is clear that the
description of charged fluids as lattice gases or as charged hard
spheres gives very different phenomenology in high field regions. The
lattice gas crosses over very rapidly to a close packed system,
whereas much higher fields are needed to compress the hard sphere
system, leading to very slow cross-overs in $1/\psi^{1/3}$ in physical
properties such as charge density.

\section{Appendices}

\subsection{Numerical methods}
We wrote numerical codes to study the double Legendre transformed free
energy
\begin{equation}
  f(c_1, c_2) - \mu(c_1+c_2) - \tilde \psi(z_1 c_1- z_2 c_2) 
\end{equation}
where $\tilde \psi=-\psi$.  We do this by working with the effective
coordinates
\begin{align}
  n =& (c_1 +c_2) \nonumber \\
  q =& (z_1 c_1 -z_2 c_2) 
\end{align}
So that we are interested in stationary points of the function.
\begin{equation}
  \tilde f(n,q) - \mu n -\tilde \psi q \label{eq:F}
\end{equation}
where we have expressed the free energy as a function of the two
independent coordinates, $n$ the number density and $q$ the charge
density.

We proceed by constructing an intermediate function $g_\mu(q)$ by
numerical minimisation of eq~(\ref{eq:F}), with fixed $\mu$ and $q$,
with $\tilde \psi=0$. The function $g_\mu(q)$ is then passed to the
Chebfun library \cite{Trefethen2007, Driscoll2014} which evaluates
$g_\mu(q)$ for different specific values of $q$ and builds a Chebyshev
approximant accurate to a relative accuracy of $10^{-15}$. From this
function we build the Legendre transform from $q$ to $\tilde \psi$ by
standard operations on $g_\mu$ \cite{Zia}.
\begin{equation}
  g_\mu(q) \rightarrow g_\mu'(q) \rightarrow {(g_\mu')}^{-1}(\tilde \psi)
  \rightarrow \int^{\tilde \psi} {(g_\mu')}^{-1}(\psi') d\psi' \label{eq:G}
\end{equation}
These steps are all performed by manipulation of the Chebyshev series, while maintaining close to
machine precision in the evaluations. The result is an approximant to $p(\tilde \psi)$. The last
step is to transform back to $p(\psi)$ which requires a flip in sign of the potential axes.

The functions $p(\psi)$ and $g_\mu(q)$ encode complementary
information on the physical system. We can find the equilibrium charge
density at a given potential from the relation
\begin{equation}
  q(\psi)= - \frac{dp}{d\psi}
\end{equation}
we find the potential at imposed charge density from
\begin{equation}
  \psi(q) = -\frac{dg_\mu}{dq }
\end{equation}
The non-standard signs in these relations come from the difference
between $\psi$ and $\tilde \psi$.


The question finally arrises as to how to use the numerically
determined curves for $p(\psi)$ in other external codes. Inspiration
comes from the {\CS} approximation for the pressure which is a ratio
of polynomials in the density.  Such a general form is an example of a
Pad\'e approximant that yields a high precision representation of the
function $p(\psi)$ with an approximation as a ratio of two cubic
polynomials that yields a rather good fit. Use of two quartics gives
results which are visually perfect. Thus the present functional forms
can be easily exported (this is even part of the chebfun library) to
simple, fast approximations that can be used in other simulation
codes.

Clearly these methods are completely general can be applied to even
more elaborate equations of state, extrapolated from the best virial
expansions \cite{virial}.

\subsection{Convex formulation for Poisson-Boltzmann free energies}
As an alternative to writing the Poisson-Boltzmann functional in terms
of the potential $\psi$ with the help of the function $p(\psi)$ we can
generate an equivalent convex formulation using the displacement field
$\bf D$. As shown in \cite{acm,justine} this exact transformation
requires the Legendre transfrom of the function $p(\psi)$. However, we
have already evaluated this object, it is just $g_\mu(q)$,
eq.~(\ref{eq:G}). We can thus at once conclude that the general convex
Poisson-Boltzmann function equivalent to those discussed above is
\begin{equation}
  f({\bf D}) = \frac{{\bf D}^2}{2\varepsilon} +g_\mu({\rm
    div\,} {\bf D} - \rho_e) \label{eq:dual}
\end{equation}
with $\rho_e$ the external, imposed charged density.  This form can be
particularly interesting for the numerical work when coupling to other
conformational degrees of freedom such as polymer chains or
biomolecules. While we do not have analytic expression for $g_\mu$ for
the {\CS} fluid it is again easy to generate the curve as a Chebyshev
polynomial and export them to an accurate and efficient form for use
in other codes.


\section{Acknowledgment}

A.C.M. is partially financed by the ANR grant {FSCF}.  R.P.  thanks
the hospitality of L'\'{E}cole sup\'{e}rieure de physique et de chimie
industrielles de la ville de Paris (ESPCI ParisTech) during his stay
as a visiting {\sl Joliot chair}\/ professor and acknowledges partial
support of the Slovene research agency (ARRS) through grant P1-0055.

\bibliographystyle{eplbib} \bibliography{rudip}

\end{document}